\input{aipcheck}

\documentclass[
    ,final            
  ]
  {aipproc}

\layoutstyle{8x11single}

\usepackage{bm}
\usepackage{combelow}
\usepackage{slashed}
\usepackage{tensor}
\usepackage{braket}
\usepackage{amsmath}
\usepackage{amssymb}

\usepackage{color} 

\begin{document}

\title{Massless rotating fermions inside a cylinder}


\author{Victor E. Ambru\cb{s}}{
  address={Center for Fundamental and Advanced Technical Research,\\ Romanian Academy Bd.~Mihai Viteazul 24, 300223 Timi\cb{s}oara, Romania. \footnote{Victor.Ambrus@gmail.com}}
}

\author{Elizabeth Winstanley}{
  address={School of Mathematics and Statistics, University of Sheffield,\\
Hicks Building, Hounsfield Road, Sheffield, S3 7RH, United Kingdom.  \footnote{E.Winstanley@sheffield.ac.uk}}
}

\begin{abstract}
We study rotating thermal states of a massless quantum fermion field inside a cylinder in Minkowski space-time.
Two possible boundary conditions for the fermion field on the cylinder are considered: the spectral and MIT bag boundary conditions.
If the radius of the cylinder is sufficiently small, rotating thermal expectation values are finite everywhere inside the cylinder.
We also study the Casimir divergences on the boundary.  The rotating thermal expectation values and the Casimir divergences have different properties
depending on the boundary conditions applied at the cylinder.  This is due to the local nature of the MIT bag boundary condition, while the spectral boundary condition is nonlocal.
\end{abstract}

\maketitle


\section{Introduction}
\label{sec:intro}

For quantum fields on black hole space-times, the Hartle-Hawking state represents a black hole in thermal equilibrium with a heat bath at the Hawking temperature \cite{Hartle:1976tp}.
On a non-rotating black hole space-time, the Hartle-Hawking state is well defined and regular everywhere outside the event horizon.
On a rotating black hole space-time,  the Hartle-Hawking state would represent a rotating heat bath in thermal equilibrium at the Hawking temperature.
For a quantum scalar field, such a state is ill-defined everywhere outside the event horizon \cite{Kay:1988mu}, unless the black hole is enclosed inside a reflecting boundary sufficiently close to the
event horizon \cite{Duffy:2005mz}.
On the other hand, for a
fermion
field, it is possible to define a rotating thermal equilibrium state on a rotating black hole space-time, but this state is divergent far from the black hole event horizon \cite{Casals:2012es}.

In order to understand this fundamental difference between the behaviour of bosonic and fermionic quantum fields on a rotating black hole space-time, it is instructive to study the toy model of a rigidly-rotating thermal state for a quantum field on Minkowski space-time.
Considering unbounded Minkowski space-time, rigidly-rotating thermal states are ill-defined everywhere for bosonic fields \cite{Duffy:2002ss} but can be defined for fermionic fields \cite{Ambrus:2014uqa}.
However, rotating thermal states for fermions become singular far from the axis of rotation \cite{Ambrus:2014uqa}, mimicking the behaviour of rotating thermal states on black hole space-times \cite{Casals:2012es}.
If instead we consider a quantum scalar field on Minkowski space-time bounded by a cylinder enclosing the axis of rotation, then, if the radius of the cylinder is sufficiently small,
regular rigidly-rotating thermal states
can be defined \cite{Duffy:2002ss}.

In this paper we consider a quantum fermion field inside a cylinder in Minkowski space-time, and construct rigidly-rotating thermal states.
We begin with a brief review of the modes of a quantum fermion field on unbounded rotating Minkowski space-time, and the definition and properties of
rotating vacuum and thermal states \cite{Ambrus:2014uqa}.
With the cylinder present, we consider two possible boundary conditions on the fermion field, namely the spectral \cite{Hortacsu:1980kv} and MIT bag
\cite{Chodos:1974je} boundary conditions.
We discuss some of the properties of rotating thermal expectation values when the boundary is not too far from the axis of rotation.
We also outline the behaviour of Casimir expectation values near the boundary.
Throughout we restrict attention to massless fermions.
Further details and the massive case will be discussed in a forthcoming paper \cite{art:ambrusprep}. Some preliminary results have previously appeared in
\cite{art:ambrusmg13}.

\section{Fermions on unbounded rotating Minkowski space-time}
\label{sec:unbounded}

Let $(t_{M}, \rho , \varphi _{M}, z)$ be cylindrical polar coordinates for Minkowski space.
Setting $\varphi = \varphi _{M} - \Omega t_{M}$, and $t=t_{M}$,  we obtain the metric for rotating Minkowski space-time:
\begin{equation}
  ds^2 = - \varepsilon \, dt^2 + 2\rho^2 \Omega \, dt\,d\varphi + d\rho^2 + \rho^2 d\varphi^2 + dz^2
  \label{eq:metric}
\end{equation}
where
\begin{equation}
\varepsilon = 1 -\rho^2 \Omega^2.
\end{equation}
An observer at fixed $(\rho, \varphi, z)$ rotates around the $z$-axis with angular speed $\Omega $.
When $\rho > \Omega ^{-1}$, such an observer must have speed greater than the speed of light.
The surface on which $\varepsilon =0$ and $\rho = \Omega ^{-1}$ is therefore called the {\em {speed of light surface}} (SOL).

The Dirac equation for massless fermions on the space-time (\ref{eq:metric}) takes the form
\begin{equation}
  \left[ \gamma^{{\hat {t}}} (H + \Omega J_z) - \mathbf{\gamma} \cdot \mathbf{P} \right] \psi(x) = 0,
  \label{eq:Dirac}
\end{equation}
where $H=i\partial _{t}$ is the rotating Hamiltonian, $J_{z}$ is the $z$-component of the angular momentum and ${\mathbf {P}}$ is the momentum operator.
Hatted indices represent tensor components with respect to a tetrad $e^{\mu }_{\hat{\alpha}}$. The gamma matrices
$\gamma^{\hat{\alpha}}$ are in the Dirac representation and satisfy
$\{\gamma^{\hat{\alpha}}, \gamma^{\hat{\rho}}\} = -2\eta^{\hat{\alpha}\hat{\rho}}$, while their covariant counterparts
$\gamma ^{\mu }= e^{\mu }_{{\hat {\alpha }}}\gamma ^{{\hat {\alpha }}}$ satisfy $\{\gamma^\mu, \gamma^\nu\} = -2g^{\mu\nu}$.

Mode solutions of (\ref{eq:Dirac}) with Minkowski energy $E$, $z$-axis momentum $k$, $z$-axis angular momentum $m + \tfrac{1}{2}$ and helicity $\lambda$ take the form:
\begin{equation}
  U^\lambda_{Ekm} = \frac{1}{4\pi} e^{-i{\widetilde {E}} t + ik z}
  \begin{pmatrix}
   \displaystyle \phi ^{\lambda }_{Ekm} \\
   \displaystyle \frac{2\lambda E}{\left| E \right| } \phi ^{\lambda }_{Ekm}
  \end{pmatrix}.
\label{eq:modes}
\end{equation}
The two-spinor $\phi^\lambda_{Ekm}$ introduced above is given by
\begin{equation}
 \phi ^{\lambda }_{Ekm} = e^{i(m + \tfrac{1}{2})\varphi}
  \begin{pmatrix}
   \displaystyle \sqrt{1 + \frac{2\lambda k}{p}} e^{-\frac{i}{2}\varphi} J_m(q\rho) \\
   \displaystyle 2i\lambda \sqrt{1 - \frac{2\lambda k}{p}} e^{\frac{i}{2}\varphi} J_{m+1}(q\rho)
  \end{pmatrix},
 \label{eq:phi}
\end{equation}
 where $J_{m}$ is a Bessel function of the first kind.
In (\ref{eq:modes}, \ref{eq:phi}), we have introduced ${\widetilde{E}} = E - \Omega(m + \tfrac{1}{2})$, which is the eigenvalue of the rotating Hamiltonian $H = i\partial_t$, and the momentum $p=\pm E={\sqrt {q^{2}+k^{2}}}$, where $q$ is the transverse momentum.

The definition of vacuum and thermal states for fermions on the unbounded space-time (\ref{eq:metric}) is discussed in detail in \cite{Ambrus:2014uqa}.
Constructing a vacuum state depends crucially on the choice of positive frequency for the modes (\ref{eq:modes}).
A rotating thermal state at inverse temperature $\beta $ is built up from the chosen vacuum state by thermally populating the modes (\ref{eq:modes}) using the rotating Hamiltonian $H$.
This means that the appropriate energy in the Planck factor in thermal expectation values (t.e.v.s) is ${\widetilde {E}}$.

There are two particular choices of positive frequency (and hence vacuum state) which are of interest:
\begin{itemize}
\item
Choosing $E>0$ \cite{Vilenkin:1980zv} leads to the nonrotating Minkowski vacuum.  Defining rotating thermal states relative to this vacuum leads to
t.e.v.s which have spurious temperature-independent terms \cite{Ambrus:2014uqa,Vilenkin:1980zv}.
\item
On the other hand, choosing ${\widetilde {E}}>0$ \cite{Iyer:1982ah} gives a rotating vacuum which is distinct from the nonrotating Minkowski vacuum.
Relative to this vacuum, t.e.v.s for rotating thermal states do not contain the above spurious temperature-independent terms.
\end{itemize}
For both choices of the vacuum state, the t.e.v.s are regular within the SOL
but diverge as the SOL is approached and $\varepsilon \rightarrow 0$ \cite{Ambrus:2014uqa}.

\section{Fermions on rotating Minkowski space-time with a cylindrical boundary}
\label{sec:bounded}

We now consider rotating Minkowski space-time (\ref{eq:metric}) with a cylindrical boundary at $\rho =R$, where $R\le \Omega ^{-1}$ so that the cylinder
is inside or on the SOL.
Boundary conditions must be imposed on the fermion modes (\ref{eq:modes}) in order that the rotating Hamiltonian $H$ is self-adjoint.
The Hamiltonian $H$ will be self-adjoint if, for any two solutions $\psi $ and $\chi $ of the Dirac equation (\ref{eq:Dirac}), we have
\begin{equation}
0 =  \braket{\psi, H\chi} - \braket{H \psi, \chi} = -i \int_{\partial V} d \Sigma_i \sqrt{-g}\, {\overline {\psi}} \, \gamma^i \chi
\label{eq:SA}
\end{equation}
where $\braket {\psi , \chi}$ is the Dirac inner product and $\partial V$ is the boundary of the volume $V$.
For a cylindrical boundary at $\rho = R$,  the surface integral in (\ref{eq:SA}) takes the form
\begin{equation}
 R \int_{-\infty}^\infty dz \int_0^{2\pi} d\varphi \, {\overline {\psi }} \gamma^{\hat{\rho}} \chi = 0.
 \label{eq:BC}
\end{equation}
We consider two implementations of the boundary condition (\ref{eq:BC}), namely the spectral \cite{Hortacsu:1980kv} and MIT bag \cite{Chodos:1974je}
boundary conditions.

\subsection{Spectral boundary conditions}
\label{sec:spectral}

Spectral boundary conditions \cite{Hortacsu:1980kv} are constructed by considering the Fourier transform of a fermion spinor $\psi $ with respect to the azimuthal angle $\varphi $:
\begin{equation}
 \psi = \sum_{m = -\infty}^\infty e^{i(m+\tfrac{1}{2})\varphi} (e^{-\frac{i}{2}\varphi} \psi_{m+\frac{1}{2}}^1,
 e^{\frac{i}{2}\varphi} \psi_{m+\frac{1}{2}}^2, e^{-\frac{i}{2}\varphi} \psi_{m+\frac{1}{2}}^3,
 e^{\frac{i}{2}\varphi} \psi_{m+\frac{1}{2}}^4)^T.
 \label{eq:transform}
\end{equation}
Using this notation, in order for the rotating Hamiltonian $H$ to be self-adjoint, it must be the case that
\begin{equation}
\sum_{m = -\infty}^\infty \left(
 \psi^{1\, *}_{ m + \frac{1}{2}} \chi^4_{ m + \frac{1}{2}} +
 \psi^{2\, *}_{m + \frac{1}{2}} \chi^3_{m + \frac{1}{2}} +
 \psi^{3\, *}_{m + \frac{1}{2}} \chi^2_{m + \frac{1}{2}} +
 \psi^{4\, *}_{m + \frac{1}{2}} \chi^1_{m + \frac{1}{2}} \right) =0,
 \label{eq:spectral1}
\end{equation}
for any $\psi$ and $\chi$ satisfying the Dirac equation \eqref{eq:Dirac}.
By considering the charge conjugate $\psi _{c}$ of the spinor $\psi $, a second condition for self-adjointness of the Hamiltonian is found:
\begin{equation}
\sum_{m = -\infty}^\infty \left(
 \psi^{1}_{ m + \frac{1}{2}} \chi^1_{ -m - \frac{1}{2}} -
 \psi^{2}_{m + \frac{1}{2}} \chi^2_{-m - \frac{1}{2}} -
 \psi^{3}_{m + \frac{1}{2}} \chi^3_{-m - \frac{1}{2}} +
 \psi^{4}_{m + \frac{1}{2}} \chi^4_{-m - \frac{1}{2}} \right) =0.
\label{eq:spectral2}
\end{equation}
To satisfy the two equations (\ref{eq:spectral1}, \ref{eq:spectral2}), we set
$\psi_{m + \frac{1}{2}}^1\rfloor_{\rho = R} =\psi_{m + \frac{1}{2}}^3\rfloor_{\rho = R} = 0$ if
$m+\tfrac{1}{2}>0$ and $\psi_{m + \frac{1}{2}}^2\rfloor_{\rho = R} =
 \psi_{m + \frac{1}{2}}^4\rfloor_{\rho = R} = 0$ if $m+\tfrac {1}{2}<0$ \cite{Hortacsu:1980kv}.
 This means that the transverse momentum $q$ is quantized so that
\begin{equation}
J_{m}\left( q_{m}R \right) =0 \quad {\mbox {if $m+\tfrac {1}{2}>0$}},
\qquad \qquad
J_{m+1}\left( q_{m}R \right) =0 \quad {\mbox {if $m+\tfrac {1}{2}<0$}}.
\label{eq:BCspectral}
\end{equation}
Due to the application of the Fourier transform in the above derivation, the spectral boundary conditions are nonlocal in the sense that the integrand in (\ref{eq:BC}) does not vanish at each point on the boundary, but the integral in (\ref{eq:BC}) is zero.
The spectral boundary conditions can be applied separately to modes of positive and negative helicity.

Using known properties of the zeros of the Bessel functions $J_{m}$ \cite{Watson}, it can be shown that, for transverse momenta $q_{m}$ satisfying (\ref{eq:BCspectral}), we have $q_{m}R>m+\tfrac {1}{2}$. Hence, for $E>0$,
\begin{equation}
ER \ge q_{m}R > m+\tfrac {1}{2} \qquad {\mbox {and}} \qquad {\widetilde {E}}R > \left( 1 - \Omega R \right) \left( m + \tfrac {1}{2} \right) .
\end{equation}
Therefore $E{\widetilde {E}}>0$ for all values of the quantum numbers $k$ and $m$ as long as the boundary is inside or on the SOL ($\Omega R \le 1$).

\subsection{MIT bag boundary conditions}
\label{sec:MIT}

In the MIT bag model \cite{Chodos:1974je}, the boundary condition (\ref{eq:BC}) is satisfied by ensuring that the integrand vanishes at any point $x_C$ on the boundary. This is achieved by setting
\begin{equation}
 i n_\mu \gamma^\mu \psi\rfloor_{x = x_C} = \psi \rfloor_{x = x_C} ,
 \label{eq:MIT}
\end{equation}
where $n_\mu $ is the normal to the boundary.
Unlike the spectral boundary condition, the MIT bag boundary condition (\ref{eq:MIT}) does not commute with the helicity operator and so the spinor $\psi $ must be a linear combination of positive and negative helicity modes:
\begin{equation}
\psi = b^+_{Ekm} U^+_{Ek m} + b^-_{Ekm} U^-_{Ek m}
\end{equation}
where the $b^{\pm }_{Ekm}$ are constants and the spinors $U^{\lambda }_{Ekm}$ are given in (\ref{eq:modes}).
The boundary condition (\ref{eq:MIT}) then reduces to a pair of linear equations for the constants $b^{\pm }_{Ekm}$ which have a nontrivial solution only if
\begin{equation}
J_{m}\left( q_{m}R \right) = \pm J_{m+1}\left( q_{m}R \right) .
\label{eq:BCMIT}
\end{equation}

Properties of the solutions $q_{m}R$ of (\ref{eq:BCMIT}) are studied in \cite{Beneventano:2014ksa}.  Using Theorems 3.1 and 3.2 from \cite{Beneventano:2014ksa}, together with known properties of the zeros of the Bessel functions $J_{m}$ and their derivatives $J_{m}'$ \cite{Watson},
it can be shown that $q_{m}R>m+\tfrac {1}{2}$ for transverse momenta $q_{m}$ satisfying (\ref{eq:BCMIT}) \cite{art:ambrusprep}.
Hence, as in the spectral case, we have $E{\widetilde {E}}>0$ for all values of the quantum numbers $k$ and $m$ as long as the boundary is inside or on the SOL ($\Omega R \le 1$).

\section{Rotating thermal expectation values}
\label{sec:thermal}

For both spectral and MIT bag boundary conditions, the fact that $E{\widetilde {E}}>0$ for all field modes means that the possible choices of positive frequency (namely $E>0$ or ${\widetilde {E}}>0$) are the same when the boundary is inside or on the SOL.
Therefore, in this case the Vilenkin \cite{Vilenkin:1980zv} and Iyer \cite{Iyer:1982ah} quantizations are equivalent, or, in other words, the rotating and nonrotating vacua coincide.

Rotating thermal states at inverse temperature $\beta $ are defined relative to this vacuum state by thermally populating the field modes satifying either the spectral or MIT bag boundary conditions, using the rotating Hamiltonian $H$.
We are interested in t.e.v.s of the component of the neutrino charge current $\braket{J^{z}}$ (all other components of the neutrino charge current vanish) and the components of the stress-energy tensor $\braket {T_{\mu \nu }}$.
Expressions for these expectation values written as mode sums are rather lengthy (particularly for the stress-energy tensor) and will be presented elsewhere,  together with detailed derivations \cite{art:ambrusprep}.
Here we present a selection of results for the spectral and MIT bag boundary conditions.

\begin{figure}[p]
\begin{tabular}{cc}
\includegraphics[width=.45\columnwidth]{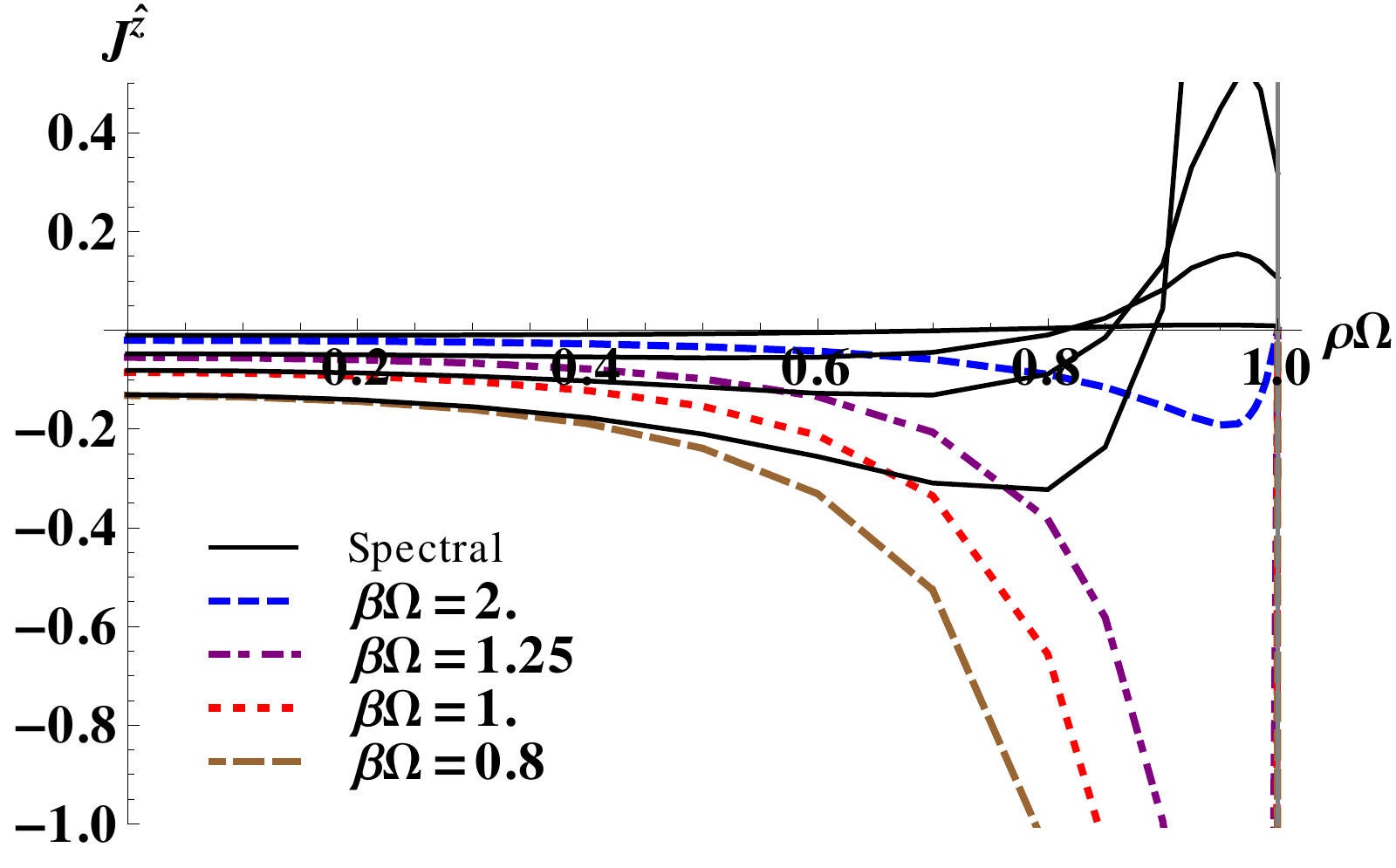} \hspace{.04\columnwidth}&
\hspace{.04\columnwidth}\includegraphics[width=.45\columnwidth]{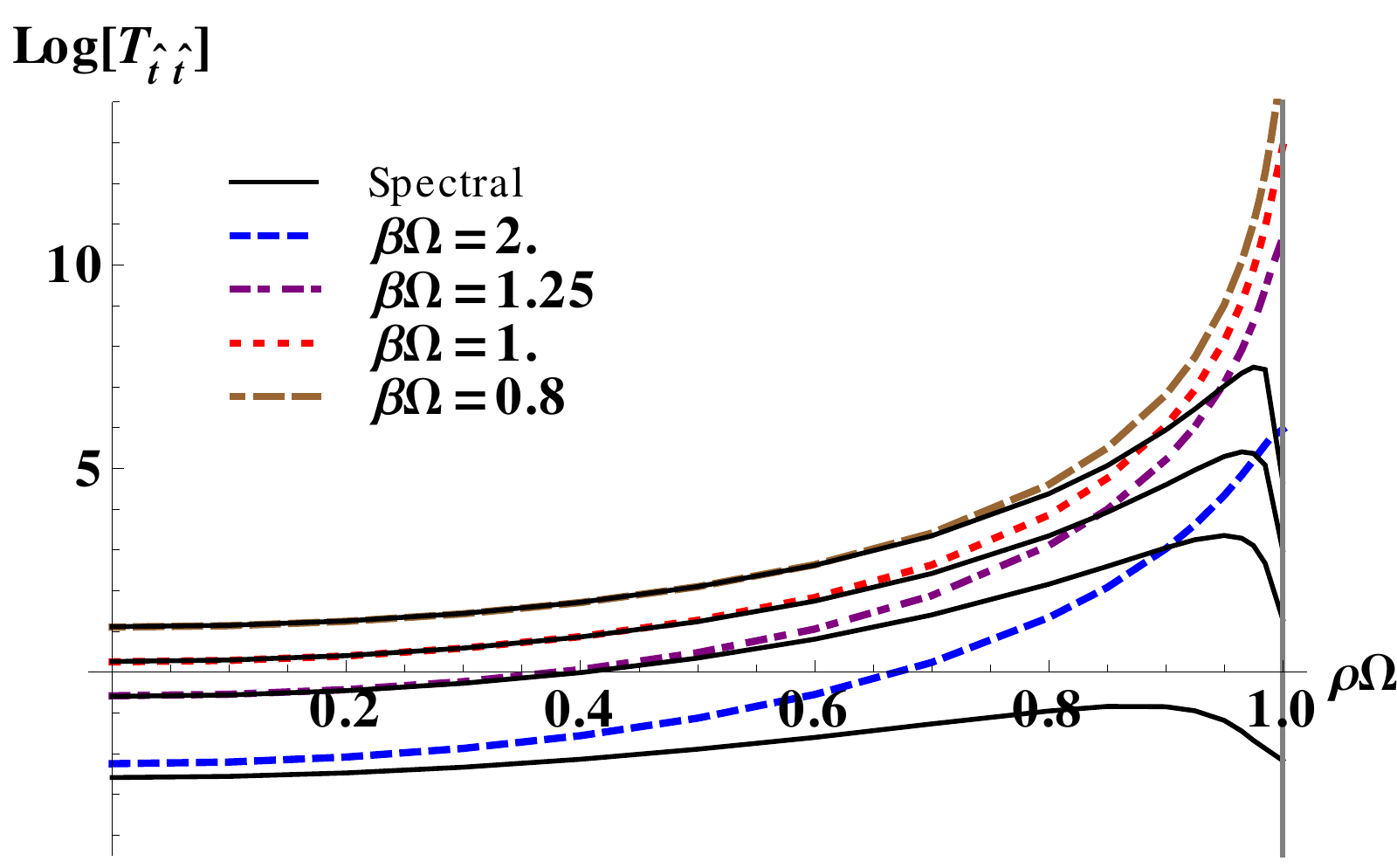}\\
\includegraphics[width=.45\columnwidth]{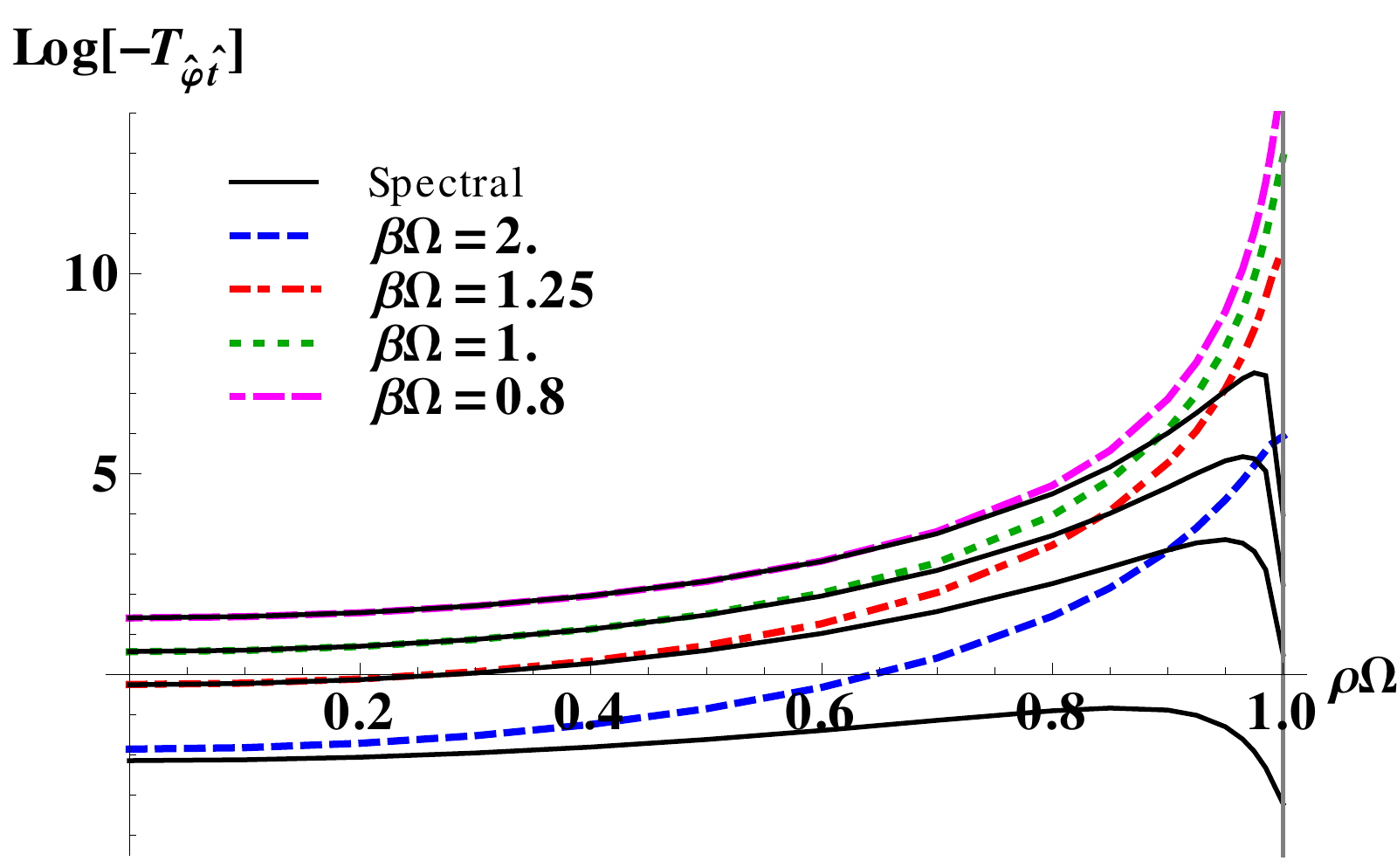} \hspace{.04\columnwidth}&
\hspace{.04\columnwidth}\includegraphics[width=.45\columnwidth]{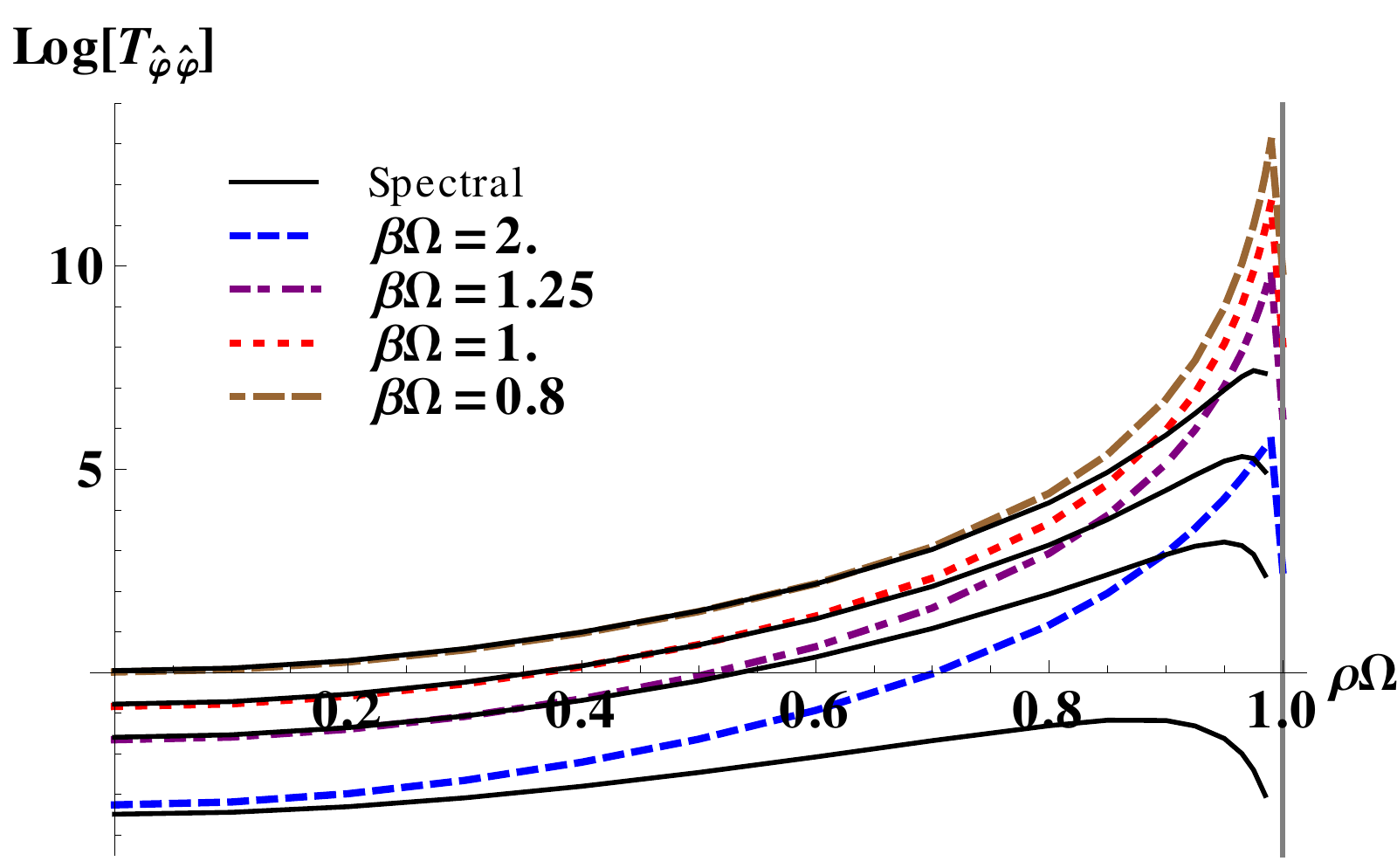}\\
\end{tabular}
\caption{Rotating t.e.v.s for massless fermions satisfying spectral (thin black lines)
and MIT bag (thick dashed coloured lines) boundary conditions as functions of the distance from the rotation axis.  The boundary is at $\rho \Omega = R\Omega =1$. The graphs show the t.e.v.s of the component of the neutrino charge current $\braket{J^{\hat {z}}}$
and the logarithms of some components of the stress-energy tensor $\braket{T_{{\hat {\mu }} {\hat {\nu }}}}$, relative to a basis of tetrad vectors \cite{art:ambrusprep}. Results are shown for a selection of values of the inverse temperature $\beta $. See Table~\ref{tab:1} for a summary of the behaviour seen in these t.e.v.s. }
\label{fig:1}
\end{figure}

\begin{table}[p]
\begin{tabular}{|p{.05\linewidth}|p{.35\linewidth}|p{.35\linewidth}|}
\hline
& Spectral & MIT\\
\hline
$J^{\hat{z}}$ & Negative on axis and positive on boundary & Vanishes as $\rho \rightarrow R$\\
$T_{{\hat{t}}{\hat{t}}}$ & Decreasing close to boundary & Increasing close to boundary \\
$T_{{\hat{\varphi}}{\hat{t}}}$ & Decreasing close to boundary & Increasing close to boundary\\
$T_{{\hat{\varphi}}{\hat{\varphi}}}$ & Vanishes as $\rho \rightarrow R$ & Decreasing close to boundary\\
\hline
\end{tabular}
\caption{Summary of the properties of the t.e.v.s shown in Figure~\ref{fig:1}.}
\label{tab:1}
\end{table}

In Figure~\ref{fig:1} we show rotating t.e.v.s for massless fermions satisfying the spectral and MIT bag boundary conditions.
The boundary is on the SOL. All t.e.v.s are finite everywhere inside and on the boundary, for both spectral and MIT bag boundary conditions.
Figure~\ref{fig:1} also shows a number of differences in the properties of t.e.v.s with different boundary conditions applied, and these properties are summarized in Table~\ref{tab:1}.

\begin{figure}[p]
\begin{tabular}{cc}
\includegraphics[width=.45\columnwidth]{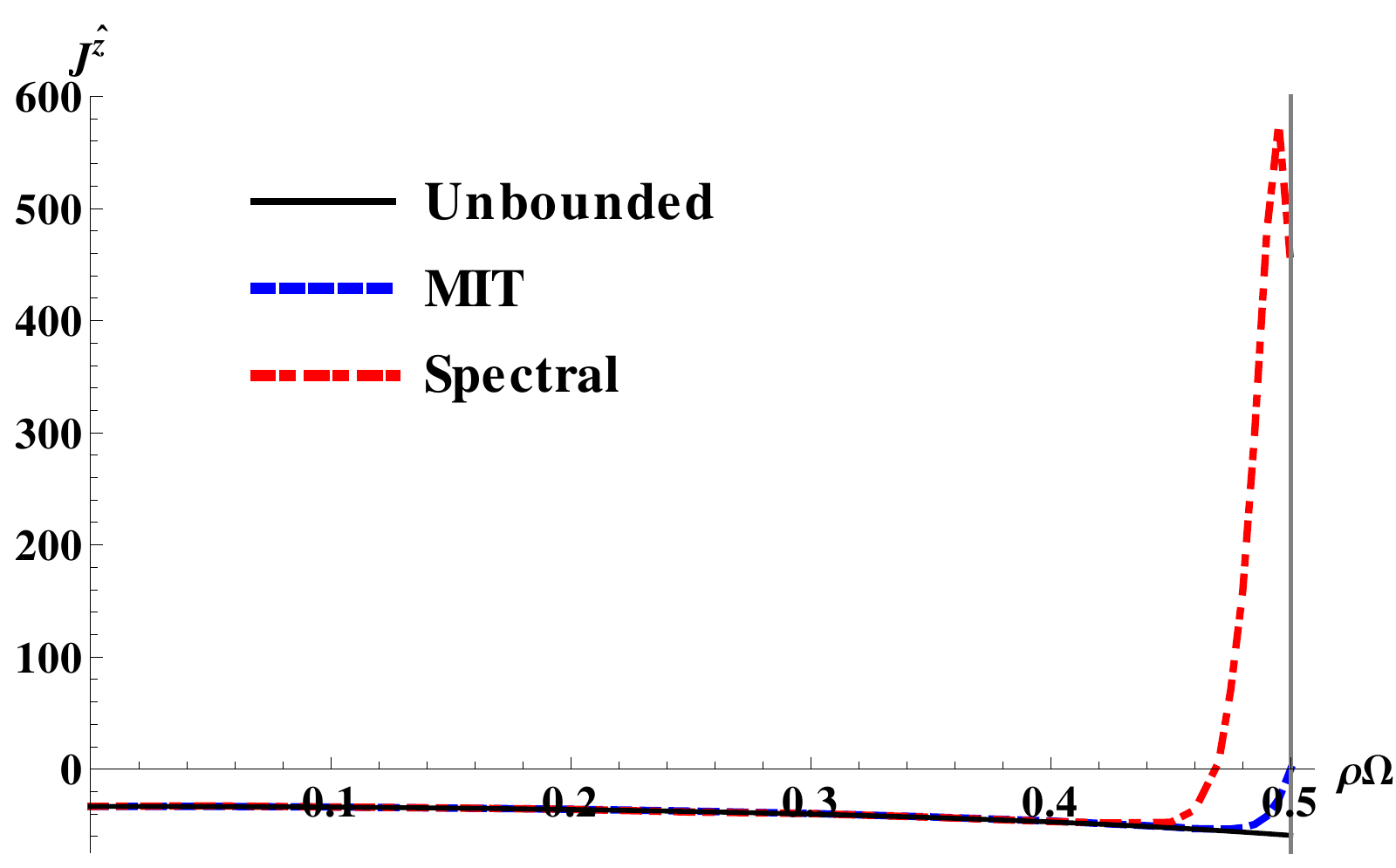} \hspace{.04\columnwidth}&
\hspace{.04\columnwidth}\includegraphics[width=.45\columnwidth]{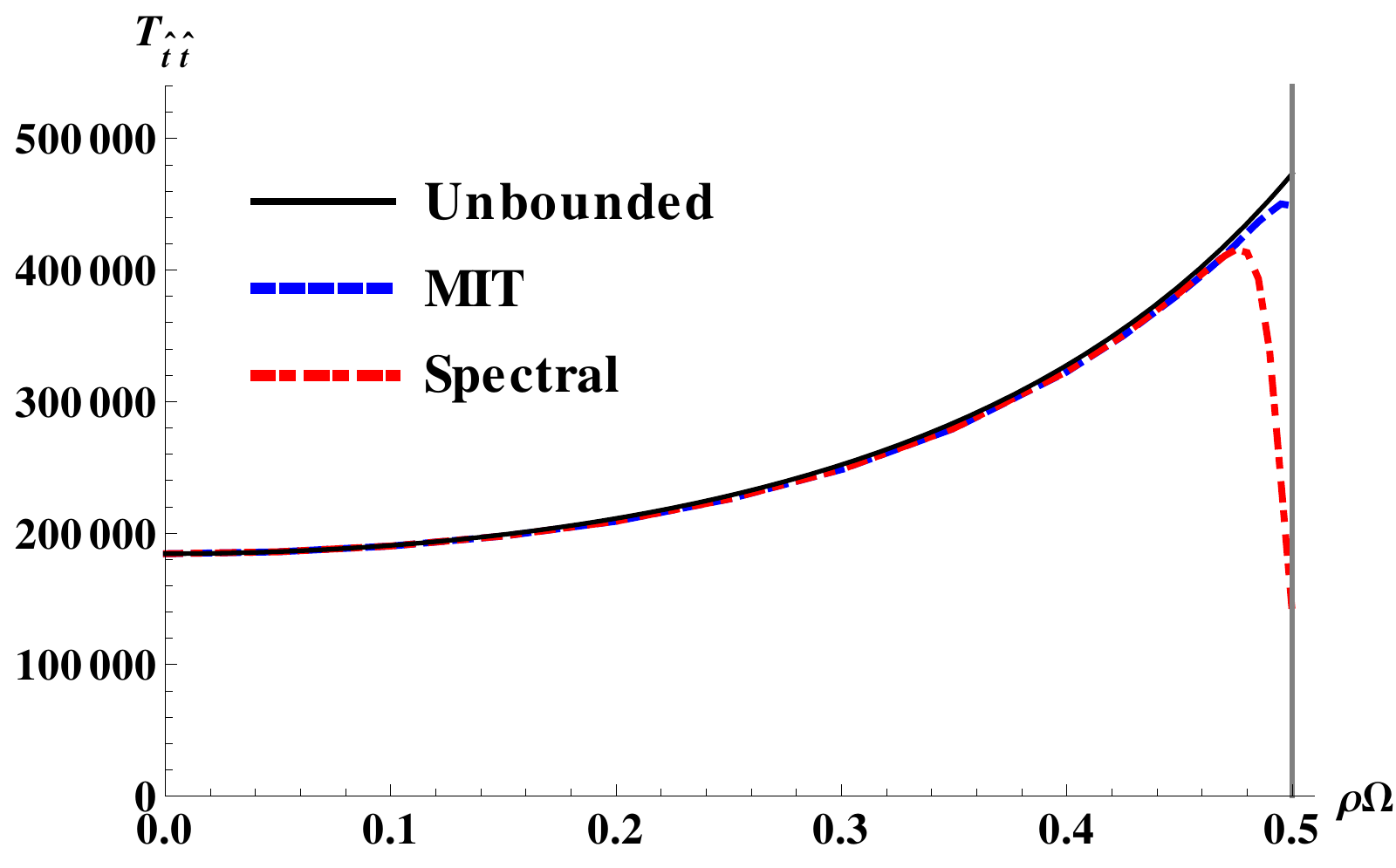}\\
\end{tabular}
\caption{Rotating t.e.v.s for massless fermions satisfying spectral (purple dotted lines) and MIT bag (blue dashed lines) boundary conditions compared with those for
massless fermions on unbounded Minkowski space-time (thin black lines). The boundary is at $R\Omega = 0.5$ and the inverse temperature is $\beta \Omega = 0.05$. }
\label{fig:2}
\end{figure}

In Figure~\ref{fig:2} we show rotating t.e.v.s when the boundary is inside the SOL and the temperature is high, comparing the t.e.v.s for the spectral and MIT bag boundary conditions with those for the unbounded rotating space-time \cite{Ambrus:2014uqa}.
It can be seen that, away from the boundary, the rotating t.e.v.s with both the spectral and MIT bag boundary conditions follow closely the profiles for unbounded rotating Minkowski space-time.
However, there are significant differences in behaviour near the boundary.
The neutrino charge current vanishes on the boundary in the MIT bag case, and is positive on the boundary for spectral boundary conditions.
The energy density $\braket{T_{{\hat {t}}{\hat {t}}}}$ near the boundary is
lower in the bounded case compared with the unbounded space-time.
This drop in energy density is particularly marked in the case of spectral boundary conditions.

\section{Casimir effect}
\label{sec:casimir}

The results in Figure~\ref{fig:2} show significant differences in rotating t.e.v.s near the boundary depending on the boundary conditions applied (or no boundary conditions in the case of the unbounded space-time).
To explore these boundary effects further, we close this paper by considering Casimir expectation values $\braket{T_{{\hat {t}}{\hat {t}}}}_{{\text {Cas}}}$,
defined as the difference between {\emph {vacuum}} expectation values for the nonrotating vacuum states on the bounded and unbounded Minkowski space-times.
This difference in vacuum expectation values diverges as the boundary is approached.
The leading-order behaviour of this divergence can be found analytically using asymptotic analysis \cite{art:ambrusprep}.
We find the following expressions, for spectral and MIT bag boundary conditions respectively  (see \cite{BezerradeMello:2008zd}
for the MIT bag result):
\begin{equation}
 \braket{T_{{\hat{t}}{\hat{t}}}}_{\text{Cas}}^{\text{spec}} \sim \frac{1}{16\pi^2 R^4 \delta ^4}, \qquad
  \braket{T_{{\hat{t}}{\hat{t}}}}_{\text{Cas}}^{\text{MIT}} \sim
 \frac{1}{120\pi^2R^4\delta ^3},
 \label{eq:Casimir}
\end{equation}
where $\delta = 1 - \frac{\rho}{R}$. 
Both expectation values in (\ref{eq:Casimir}) diverge as
$\delta \rightarrow 0$
and the boundary is approached.
The $\delta ^{-3}$ divergence for the MIT bag boundary conditions agrees with that predicted by Deutsch and Candelas \cite{Deutsch:1978sc} for local boundary conditions, in accordance with the local nature of the MIT bag boundary conditions.
However, for spectral boundary conditions, the
divergence is one order of magnitude more severe. This is not in contradiction with the analysis of
\cite{Deutsch:1978sc}, since the latter
is not valid for the nonlocal spectral boundary conditions.

\section{Conclusions}
\label{sec:conc}

In this paper, we have considered a massless quantum fermion field on Minkowski space-time bounded by a cylinder of radius $R$. In particular, we have defined vacuum and thermal states which are rigidly rotating with angular speed $\Omega $ about the $z$-axis.
We focused on the case $\Omega R\le 1$, for which the boundary is inside or on the speed of light surface.
The rotating vacuum is then identical to the nonrotating vacuum and thermal expectation values are finite everywhere inside and on the boundary.
We considered two boundary conditions for the fermion field on the cylinder, namely the spectral and MIT bag boundary conditions.
The MIT bag boundary conditions are local in nature, but the spectral boundary conditions are nonlocal as they are derived from a Fourier transform of the fermion field.
This difference between the boundary conditions is manifest both in different behaviour of the thermal expectation values as the boundary is approached, and in the rate of divergence of Casimir expectation values.


\begin{theacknowledgments}
V.E.A.~was supported by a studentship from the School of Mathematics and Statistics at the University of Sheffield.
The work of E.W.~is supported by the Lancaster-Manchester-Sheffield Consortium for
Fundamental Physics under STFC grant ST/L000520/1.
\end{theacknowledgments}

\bibliographystyle{aipproc}   

\begin{thebibliography}{99}

\bibitem{Hartle:1976tp}
  J.~B.~Hartle and S.~W.~Hawking,
  \emph {Phys.\ Rev.\ D} {\textbf {13}}, 2188--2203 (1976).

\bibitem{Kay:1988mu}
  B.~S.~Kay and R.~M.~Wald,
  {\emph {Phys.\ Rept.}} {\textbf {207}}, 49--136 (1991).

\bibitem{Duffy:2005mz}
  G.~Duffy and A.~C.~Ottewill,
  {\emph {Phys.\ Rev.\ D}} {\textbf {77}}, 024007 (2008).

\bibitem{Casals:2012es}
  M.~Casals, S.~R.~Dolan, B.~C.~Nolan, A.~C.~Ottewill and E.~Winstanley,
  {\emph {Phys.\ Rev.\ D}} {\textbf {87}},  064027 (2013).

\bibitem{Duffy:2002ss}
  G.~Duffy and A.~C.~Ottewill,
  \emph{Phys.\ Rev.\ D} {\textbf {67}}, 044002 (2003).

\bibitem{Ambrus:2014uqa}
  V.~E.~Ambru\cb{s} and E.~Winstanley,
  \emph{Phys.\ Lett.\ B} {\textbf {734}}, 296--301 (2014).

\bibitem{Hortacsu:1980kv}
M.~Horta\c{c}su,
  K.~D.~Rothe and B.~Schroer,
  {\emph {Nucl.\ Phys.\ B}} {\textbf {171}}, 530--542 (1980).

\bibitem{Chodos:1974je}
  A.~Chodos, R.~L.~Jaffe, K.~Johnson, C.~B.~Thorn and V.~F.~Weisskopf,
  {\emph {Phys.\ Rev.\ D}} {\textbf {9}}, 3471--3495 (1974).

\bibitem{art:ambrusprep}
V.~E.~Ambru\cb{s} and E.~Winstanley,  \emph{Thermal states for rotating fermions inside a cylindrical boundary},
paper in preparation.

\bibitem{art:ambrusmg13}
V.~E.~Ambru\cb{s} and E.~Winstanley, \emph{Proceedings of the thirteenth Marcel Grossman meeting MG13},
{\tt {arXiv:1302.3791 [gr-qc]}}.

\bibitem{Vilenkin:1980zv}
  A.~Vilenkin,
  {\emph {Phys.\ Rev.\ D}} {\textbf {21}}, 2260--2269 (1980).

\bibitem{Iyer:1982ah}
  B.~R.~Iyer,
  {\emph {Phys.\ Rev.\ D}} {\textbf {26}}, 1900--1905 (1982).

\bibitem{Watson}
G.~N.~Watson, {\emph {A treatise on the theory of Bessel functions}}, Cambridge University Press (1922).

\bibitem{Beneventano:2014ksa}
  C.~G.~Beneventano, I.~V.~Fialkovsky and E.~M.~Santangelo,
  \emph {Zeroes of combinations of Bessel functions and mean charge of graphene nanodots},
  {\tt {arXiv:1407.0615 [math-ph]}}.

\bibitem{BezerradeMello:2008zd}
  E.~R.~Bezerra de Mello, V.~B.~Bezerra, A.~A.~Saharian and A.~S.~Tarloyan,
 {\emph { Phys.\ Rev.\ D}} {\textbf {78}}, 105007 (2008).

\bibitem{Deutsch:1978sc}
  D.~Deutsch and P.~Candelas,
  {\emph {Phys.\ Rev.\ D }} {\textbf {20}}, 3063--3080 (1979).


\end{thebibliography}


\end{document}